\documentclass[lettersize,journal]{IEEEtran}
\usepackage{amsmath,amsfonts}
\usepackage{algorithmic}
\usepackage{algorithm}
\usepackage{array}
\usepackage[caption=false,font=normalsize,labelfont=sf,textfont=sf]{subfig}
\usepackage{textcomp}
\usepackage{stfloats}
\usepackage{url}
\usepackage{verbatim}
\usepackage{graphicx}
\usepackage{cite}
\usepackage{longtable}
\usepackage{makecell}
\usepackage{multicol}
\usepackage{multirow}
\UseRawInputEncoding
\begin{document}

\title{Insights on the Next Generation WLAN: High Experiences (HEX)}

\author{Authors
\thanks{XXX are with School of Electronics and Information, Northwestern Polytechnical University, Xi'an, Shaanxi, China.}
}
\author{Mao~Yang,~\IEEEmembership{Member,~IEEE,}
        Bo~Li,~\IEEEmembership{Member,~IEEE,}
        Zhongjiang~Yan,~\IEEEmembership{Member,~IEEE,}

\thanks{M. Yang, Z. Yan, B. Li, and Q. Li are with School of Electronics and Information, Northwestern Polytechnical University, Xi'an, Shaanxi, China.}
}

\markboth{Journal}%
{Shell \MakeLowercase{\textit{et al.}}: A Sample Article Using IEEEtran.cls for IEEE Journals}


\maketitle

\begin{abstract}
Wireless local area network (WLAN) witnesses a very fast growth in the past 20 years by taking the maximum throughput as the key technical objective. However, the quality of experience (QoE) is the most important concern of wireless network users. In this article, we point out that poor QoE is the most challenging problem of the current WLAN, and further analyze the key technical problems that cause the poor QoE of WLAN, including fully distributed networking architecture, chaotic random access, awkward ``high capability'', coarse-grained QoS architecture, ubiquitous and complicated interference, ``no place'' for artificial intelligence (AI), and heavy burden of standard evolving. To the best of our knowledge, this is the first work to point out that poor QoE is the most challenging problem of the current WLAN, and the first work to systematically analyze the technical problems that cause the poor QoE of WLAN. We highly suggest that achieving high experiences (HEX) be the key objective of the next generation WLAN.
\end{abstract}

\begin{IEEEkeywords}
Wireless Local Area Network, IEEE 802.11, Quality of Experiences, WiFi, 802.11be, Ultra High Reliability.
\end{IEEEkeywords}

\section{Introduction}
\IEEEPARstart{W}{ireless} network becomes one indispensable and fast-growing technology for human lives. After more than 20 years of development, wireless local area network (WLAN) as well as cellular network has become one dominant type of wireless network. WLAN is standardized by IEEE 802.11, and the IEEE 802.11ax is the latest commercially available WLAN standard. The IEEE 802.11ax, also called high efficiency (HE) or WiFi 6, was officially released in 2021 \cite{11axSurvey,QuSurvey,11ax_OFDMA_survey}. Now, the industry and academia are focusing on the key technology research and standardization of IEEE 802.11be, which is the next generation of IEEE 802.11ax. The IEEE 802.11be, also called extremely high throughput (EHT) or WiFi 7, is expected to be officially released in 2024 or 2025 \cite{11be_survey2,11be_survey,11be_survey3}. It is worth noting that the IEEE 802 Standard Committee established the ultra-high reliability (UHR) study group (UHR SG) that is supposed to be the next generation WLAN standard beyond IEEE 802.11be in September 2022. It can be seen that WLAN technology and its standardization process are developing very rapidly.

Let's analyze the development of IEEE 802.11. Up to UHR that is supposed to be WiFi 8, IEEE 802.11 has gone through and is going through eight major versions including IEEE 802.11a,b,g,n,ac,ax,be, and UHR. Except for IEEE 802.11ax, whose technical objective is high efficiency, other WLAN standards take the maximum throughput as the key technical objective. Of course, throughput is quite important because higher throughput means the WLAN can provide more capacity for the wireless traffic. However, with the increasing diversity of the wireless services such as virtual reality, meta universe, ultra-high resolution online video, real-time game, remote medical, and industry applications, it is increasingly difficult for WLAN to guarantee the quality of services (QoS) of these diverse services. More importantly, the quality of experience (QoE) is the most important concern of wireless network users. The QoE is not only related to the QoS, but also related to more other factors such as human subjective factors. This makes the QoE an important and urgent factor for wireless network.

Poor quality of experience (QoE) is the most challenging problem of the current WLAN. In China, in many places, as long as the traffic usage fee of the cellular network is within the budget, people including us would rather choose 4G/5G than FREE WiFi. The problem is not the peak performance, but the stability of performance. WiFi performance changes dramatically over time. For example, in my lecture, my students told me that almost the only reason to use WiFi was that it was cheap rather than experiences. In practice, the claimed ``very/extremely high throughput'' can hardly be experienced. Therefore, we believe the high experiences or named high QoE should be the key objective of the next generation WLAN standard. In this article, we analyze the key technical problems that cause the poor QoE of WLAN including fully distributed networking architecture, chaotic random access, awkward ``high capability'', coarse-grained QoS architecture, ubiquitous and complicated interference, ``no place'' for artificial intelligence (AI), and heavy burden of standard evolving. To the best of our knowledge, this is the first work to point out that poor QoE is the most challenging problem of the current WLAN, and the first work to systematically analyze the technical problems that cause the poor QoE of WLAN.

\section{Overview of Key Technical Problems}

\begin{figure*}[!t]
\centering
\includegraphics[width=7in]{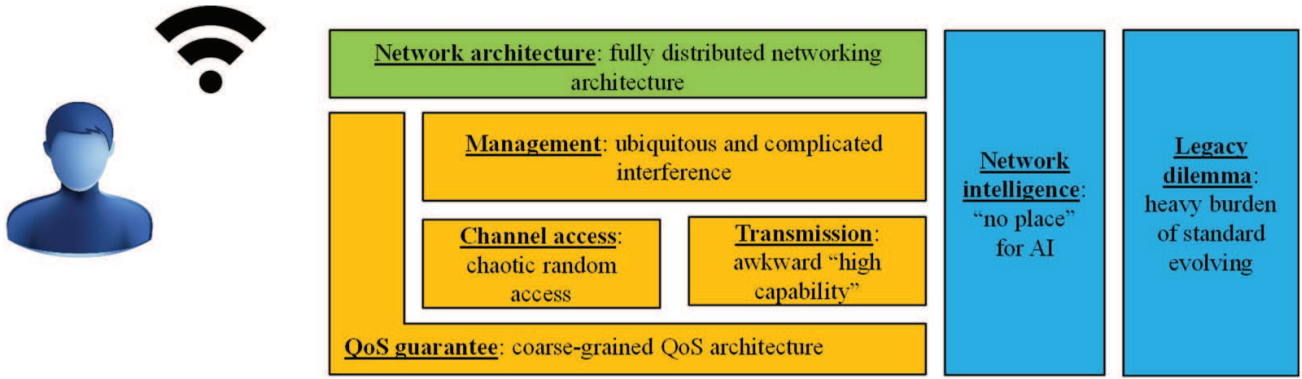}
\caption{Architecture of the key technical problems that cause the poor QoE of WLAN.}
\label{fig_architecture}
\end{figure*}

Based on our analysis, there are seven key technical problems that cause the poor QoE of WLAN. They are summarized as follows:
\begin{itemize}
  \item Network architecture: fully distributed networking architecture. Different from cellular network, WLAN adopts fully distributed network architecture. Distributed architecture has its advantages, but it usually leads to unordered network status, resulting in low QoE.
  \item Channel access: chaotic random access. The STAs including APs randomly contend the channel, resulting in chaotic channel access between STAs.
  \item Transmission: awkward ``high capability''. There are lots of new features for IEEE 802.11be such as multiple link operation (MLO), larger bandwidth such as 320MHz, higher-order modulation such as 4096-QAM, 1024-aggregation, more spatial streams (16ss), and etc. These potential features sound attractive. However, the actual available performance is far from the capability.
  \item QoS guarantee: coarse-grained QoS architecture. The current QoS architecture is traffic class based. Eight traffic identifiers (TIDs) and four access categories (ACs) cannot keep pace with the increasingly diverse services.
  \item Interference management: ubiquitous and complicated interference. Quite different from cellular network, the random interferences from overlapping basic service set (OBSS), intra-BSS, and non-WiFi system are ubiquitous and complicated.
  \item Network intelligence: ``no place'' for artificial intelligence (AI). User experience is subjective and complicated. Make network more intelligent is a natural and promising target, but it seems there is no place for AI in standardization.
  \item Legacy dilemma: heavy burden of standard evolving. It seems that every time the IEEE 802.11 standard version evolving has no choice but to coexist with many versions of legacy STAs.
\end{itemize}

As shown in Fig. \ref{fig_architecture} shows, network architecture is the framework problem derived from topology, deployment and composition. Channel access problem and transmission problem are summarized from the perspective of a single BSS or any BSS, while the interference management problem is analyzed from the perspective of multiple BSS or named OBSS. Furthermore, the QoS guarantee problem is obtained by analyzing the problems existing in the QoS guarantee methods at all levels including channel access, transmission, and interference management. Moreover, the network intelligence problem and the legacy dilemma problem are two relatively independent aspects but have a significant impact on WLAN performance and standardization. In the following section, we analyze these technical problems in detail.

\section{Analysis of Key Technical Problems}
\subsection{Network architecture: fully distributed networking architecture}
Different from cellular network, WLAN adopts fully distributed network architecture. Distributed network architecture means there is no central controller to control and manage the behaviors of multiple BSSs. Not only that, even in one single BSS, the the control function and management function of the AP ares weak as well. Normally, the AP and STAs independently make their own decisions without coordination. Distributed architecture has its advantages such as flexibility and easy deploy, but it usually leads to unordered network status, resulting in low QoE. Therefore, fully distributed networking architecture makes the network low efficiency, resulting in low QoE.

For example, within one BSS, the network is distributed although we have an AP. Both the AP and the STAs randomly contend for channel access. The STAs can determine their own configurations and parameters for channel access, transmission, and etc.. Furthermore, the coordination within extended service set (ESS) is still very limited. Even in deployed enterprise network, distributed feature also results in many problems such as starvation problem (flow-in-the-middle) which makes it quite difficult for some BSSs or STAs to access the channel.

The impact factors of the fully distributed networking architecture are summarized as follows.
\begin{itemize}
  \item Lack of overall and top-level architecture design. Since the birth of IEEE 802.11, WLAN has always followed the distributed networking architecture in order to facilitate deployment. With the standard versions evolving from generation to generation, this distributed architecture is more deeply rooted and difficult to change.
  \item Fully distributed networking architecture for network management, control, and data transmission. This means that all aspects of the network functions are designed and developed on the basis of the distributed architecture, which causes the whole system to be affected by the one.
  \item The current industrial, scientific, medical (ISM) band is limited and the environment is complicated. WLAN operates on the ISM band such as 2.4GHz, 5GHz, and 6GHz. The frequency bandwidth of ISM band is limited, which leads to insufficient resources for WLAN architecture changes. Moreover, there are many wireless network types work on the ISM band such as WiFi, Bluetooth, Zigbee, and microwave oven. Thus, the channel environment is quite complicated, which increases the technical and policy difficulties for WLAN to change the distributed architecture.
\end{itemize}

\subsection{Channel access: chaotic random access}
In the media access control (MAC) layer, the STAs including APs randomly contend the channel according to the enhanced distribution channel access (EDCA) scheme based on carrier sense multiple access with collision avoid (CSMA/CA) mechanism, resulting in chaotic channel access between STAs. Chaotic random access significantly exacerbates the collision and ineffective access and transmission, which leads to low QoE.

For example, STAs radically access the channel by consistently using the maximum TX power, setting the duration as TXOP limit that is larger than actual demand, and etc. For home and enterprise scenario, multiple BSSs or multiple STAs in one BSS are also ``enemies'' of each other in many places. Moreover, IEEE 802.11be introduces restricted target wakeup time (r-TWT) to improve the latency performance, but the rule ``Non-AP EHT STAs may behave as if overlapping quiet intervals do not exist'' exacerbates random collision. Once, a clean 6GHz band is ready for WiFi, but we roughly copied the CSMA/CA mechanism directly and contaminated this frequency band.

The impact factors of the chaotic random access are summarized as follows.
\begin{itemize}
  \item Deep-rooted random channel access. For the first version of IEEE 802.11, CSMA/CA mechanism is adopted. Although the channel access mechanism has been improved during the WLAN standard evolution, for example from distributed channel function (DCF) to EDCA, the CSMA/CA mechanism based on the idea of random access is still deep-rooted.
  \item Rigid CCA and ED thresholds for diverse scenarios. In order to support CSMA/CA, IEEE 802.11 adopts energy detection (ED) threshold and clear channel assessment (CCA) threshold to judge the channel status. However, the rigid CCA and ED do not dynamically change with the environment status varies.
  \item Lack of clean designed channel access structure. Clean designed channel access mechanisms such as scheduled channel access and reservation based channel access have several performance advantages. These channel access mechanisms can be important supplements for random channel access, and further to form a more complete channel access structure.
  \item The channel access rule for new band. Once, a clean 6GHz band is ready for WiFi, but we roughly copied the CSMA/CA mechanism directly and contaminated this frequency band. If there are new bands for WLAN in the future, can we fully seize the opportunity to carry out new design.
\end{itemize}

\subsection{Transmission: awkward ``high capability''}
There are lots of new features for IEEE 802.11be such as multiple link operation (MLO), larger bandwidth such as 320MHz, higher-order modulation such as 4096-QAM, 1024-aggregation, more spatial streams (16ss), and etc. These potential features sound attractive. However, the actual available performance is far from the capability. A series of valuable technologies cannot be achieved in practice, thus high experiences are also unbelievable.

For example, in the high-dense deployment scenario, the complicated interference among multiple BSSs and STAs make it nearly impossible to achieve 320MHz bandwidth. Maybe 20M or 40MHz is normal. Furthermore, the 4096-QAM, even 1024-QAM, is difficult to use because of the unbelievable signal TO interference and noise ratio (SINR) threshold. Remember that, 5G only uses 256-QAM as the highest modulation order. 16ss is also challenging because of the inter-stream interference and complicated user grouping.

The impact factors of the awkward ``high capability'' are summarized as follows.
\begin{itemize}
  \item Complicated channel environment and interference. The WLAN operates at the ISM band, resulting in more cross-system interferences naturally. The fully distributed architecture and the chaos random access significantly increase the inter-BSS and inter-STA interferences. In this case, the channel environment and the interference become quite complicated. Thus, the conditions of the potential high performance enabling technologies cannot meet.
  \item Lack of mechanism to make these valuable technologies work in practice. There are few mechanisms to guarantee the conditions that the potential high performance enabling technologies can meet. This makes these technologies become theoretically high performance in many cases.
  \item Increasingly low efficiency. The higher the peak throughput, the lower the efficiency. The most important reason is that higher peak throughput leads to relatively larger protocol overhead because the transmission duration of the control frames, channel access time, and the inter-frame space are not reduced along with the data transmission rate. Moreover, some rules restrict each other. For example, the maximum physical layer protocol data unit (PPDU) length will restrict the aggregation size.
\end{itemize}

\subsection{QoS guarantee: coarse-grained QoS architecture}
The current QoS architecture is traffic class based mechanism. The traffic received from upper layer are mapped into eight traffic identifiers (TIDs) and four access categories (ACs). But, the traffic (or service) type is getting increasingly diverse. Thus, the current class based QoS mechanism cannot keep pace with the increasingly diverse services. QoS is the premise of QoE, and poor QoS will inevitably lead to poor QoE.

For example, increasingly diverse services, e.g. virtual reality (VR), real-time cloud game, metaverse, digital twin, industry, remote medical, and etc., require quite diverse and challenging QoS. Furthermore, it seems we have to face to the ``unsolvable'' low latency problem. Although low latency is one feature of IEEE 802.11be, there are little effective solutions till IEEE 802.11be Draft 3.0 except r-TWT. Several standard proposals discuss less than $1ms$ latency guarantee, which seems as an ``unsolvable'' objective for WiFi. Moreover, complicated environment leads to instable resource acquisition, further resulting in unstable performance.

Internet of things (IoTs) is considered as one important scenario. However, WLAN only pursues increasingly large bandwidth and other enablers, which are ``unfriendly'' to IoTs.

Support for mobility is increasingly important for WLAN. However, seamless WiFi roaming is challenging for the current standard.

The impact factors of the coarse-grained QoS architecture are summarized as follows.
\begin{itemize}
  \item QoS architecture with appropriate granularity. The current class based QoS mechanism especially eight TIDs and four ACs cannot meet the requirements of the increasingly diverse services. Thus, other QoS mechanisms such as packet level or finer granularity should be fully studied. It is worth noting that we propose particle access method and theory for (wireless) network by treating each packet as one ``particle'' \cite{Particle_access}. For each ``particle'', we utilize the packet level resource allocation and provide the packet level QoS guarantee method to guarantee the latency and the throughput requirements of the ``particle''.
  \item Fine scalability. The QoS mechanism should keep pace with the continuously varying of wireless services. Thus, better scalability is quite important.
  \item Truly effective solution for low latency/jitter. WLAN has great challenges in ensuring low latency. If low latency/jitter cannot be guaranteed, the future of WLAN will be bleak.
  \item Stable resource acquisition. Only with stable access to resources can QoS and QoE be improved. However, there is a lack of mechanism for resource acquisition.
  \item Integrated wide band system and narrow band IoTs. The IoTs usually require small bandwidth (narrow band) in many scenarios because of the energy consumption and cost. But, large bandwidth has always been the technical goal of WLAN. For the future, if the wide band system and narrow band cannot be integrated in WLAN standard, the WLAN will probably not keep up with the development tide of the IoTs.
  \item Seamless WiFi roaming. As the operating frequency becomes higher and higher, the BSS becomes smaller and denser. So the mobility of nodes has to be considered. If the QoS cannot be guaranteed during the mobility, the QoE will be seriously affected. The challenge is that unlike cellular networks, WLAN did not take supporting mobility as an important goal at the beginning.
\end{itemize}

\subsection{Interference management: ubiquitous and complicated interference}
Quite different from cellular network, the random interferences from OBSS, intra-BSS, and non-WiFi system are ubiquitous and complicated. Interference leads to poor network performance and, consequently, poor network performance leads to poor QoE.

For example, the high-dense deployment scenario will be the typical scenario for the future wireless. WLAN devices made from different vendors adopt different implementations to select channels, bandwidth, access parameters, and etc., which makes the interferences among BSSs extremely complicated. IEEE 802.11ax introduces spatial reuse (SR), which further increases the interferences among BSSs. Moreover, the cross-system interferences may from many kinds of non-WiFi system such as licensed assisted access (LAA), LTE-u, and etc..

The impact factors of the ubiquitous and complicated interference are summarized as follows.
\begin{itemize}
  \item Lack of inter-BSS coordination. The scenarios become increasingly dense, leading to ubiquitous and complicated interference. However, the inter-BSS coordination is very limited in the current WLAN.
  \item Complicated ISM band interferences. The WLAN operates on the ISM band. There are many heterogeneous systems such as Bluetooth, LAA, LTE-u, Zigbee work on the same band. More importantly, the device cannot understand the frame sent from different system. Thus, the cross-system interferences become ubiquitous.
  \item Lack of explicit technical rules for the coexistence between WLAN and other systems. For the coexistence of the heterogeneous systems, there lacks of explicit technical rules such as channel access rules among different systems, which aggravates the complexity of interferences.
  \item Lack of new band for WLAN. New band can alleviate the interferences, but the spectrum resources are limited.
\end{itemize}

\subsection{Network intelligence: ``no place'' for AI}
User experience is subjective and complicated. Make network more intelligent is a natural and promising target for WLAN, but it seems there is ``no place'' for AI in standardization. Network intelligence can help us to evaluate the complicated user experience, and help us to choose the optimal method to guarantee the QoE. Many supervised learning (SL) and reinforcement learning (RL) w/o deep learning (DL) studies are proposed in recent years, but they lack of the supports from the standard, which makes the AI solutions difficult to extend.

For example, several standard proposals discuss the importance of AI for WLAN. But, it seems that the machine learning is usually considered as internal tools for each single module in single device. Such kind of AI obtains limited performance gain and poor scalability. Moreover, there are lots of AI models, training methods, and algorithms. Simply standardize specific one is unscalable.

To better embrace AI, we should answer following questions. What can AI do for WLAN? What can the standardization do for AI? What is the AI standardization architecture in IEEE 802.11? At least, network AI means not simply implementing specific AI model or algorithm.

\subsection{Legacy dilemma: heavy burden of standard evolving}

\begin{figure*}[!t]
\centering
\includegraphics[width=7in]{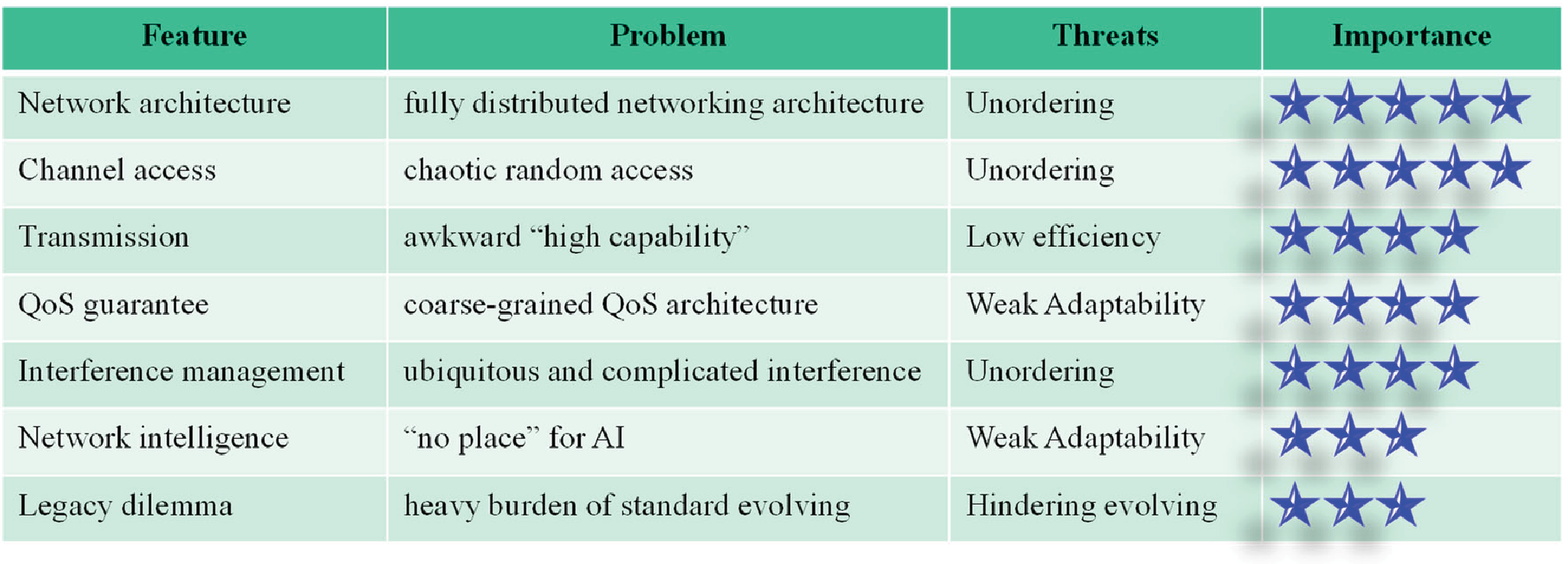}
\caption{Summary of the key technical problems that cause the poor QoE of WLAN.}
\label{fig_summary}
\end{figure*}

Different from cellular network, it seems that every time the IEEE 802.11 standard version evolving has no choice but to coexist with many versions of legacy STAs, which is a heavy burden of standard evolving. Drop the heavy burden of legacy benefits to the technical innovation because it is easy for us to focus on and further guarantee the enhancement of the user experiences of the new generation standard.

For cellular network, the standard evolving is unfettered, i.e. new band, new design, new frame format, and etc.. But, for WLAN, the standard evolving have to keep good backward compatibility, which is a ``double-edged sword''. Once, a clean 6GHz band is ready for WiFi, but we simply summoned all the old technologies. Moreover, the frame format is continuously patched. With the evolution of the standard, the patch is in tatters.

The impact factors of the heavy burden of standard evolving are summarized as follows.
\begin{itemize}
  \item Old versions and new version share the same band. This makes it difficult for the new generation of standards to break away from coexistence and carry out new design.
  \item The devices especially the APs installed with new standard have to serve all old versions. This makes it impossible for the new the standard version to break away from compatibility and carry out new design.
  \item Frame format is full of patch, but it seems that we still pay little attention to the scalable design or clean-sheet design.
\end{itemize}

\section{Conclusion}

In this article, we point out that poor QoE is the trickiest problem affecting the evolving and user population of WLAN. More importantly, we analyze seven technical problems that cause the poor QoE of WLAN. To the best of our knowledge, this is the first work to point out that poor QoE is the most challenging problem of the current WLAN, and the first work to systematically analyze the technical problems that cause the poor QoE of WLAN. The Fig. \ref{fig_summary} summarizes the key technical problems and further analyzes the threat type and the importance of each problem.

The vision of wireless network is, in our opinion, to enable people to enjoy wireless and to enrich ubiquitous interconnection. We are glad that WLAN has achieved and is achieving great success. It is the time for us to carefully think things from our god¡¯s (i.e., users¡¯) perspective. Therefore, to achieve high experiences (HEX) is highly suggested as the key objective of the next generation WLAN.


\bibliographystyle{IEEEtran}
\bibliography{mybibfile}

\newpage


\vfill

\end{document}